\documentclass[11pt]{article}

\usepackage[margin=1in]{geometry}
\usepackage[utf8]{inputenc}
\usepackage[T1]{fontenc}
\usepackage{array}
\usepackage{longtable}
\usepackage{booktabs}
\usepackage{enumitem}
\usepackage{amsmath}
\usepackage{natbib}
\usepackage{hyperref}
\usepackage{xurl}
\usepackage{xcolor}
\usepackage{parskip}
\usepackage{setspace}

\hypersetup{
  colorlinks=true,
  linkcolor=blue!50!black,
  citecolor=blue!50!black,
  urlcolor=blue!50!black
}

\newcommand{\propparskip}{\setlength{\parskip}{10pt plus 2pt minus 2pt}}

\newlength{\proptablewidth}
\newlength{\propLwidth}
\newcolumntype{L}{>{\arraybackslash\propparskip}p{\propLwidth}}
\newcolumntype{R}{>{\arraybackslash\propparskip}p{\dimexpr\proptablewidth-\propLwidth-1em\relax}}

\newenvironment{proptable}{%
  \begin{longtable}{@{}L @{\hspace{1em}} R@{}}
  \toprule
}{%
  \\ \bottomrule
  \end{longtable}
}

\newcommand{\propsep}{\\ \midrule}

\title{\textbf{Evaluating for the Long Term: Learnings from Industry}}

\author{%
\parbox{0.92\textwidth}{\centering
Leif Sigerson (Pinterest), Tom Cunningham (METR), Winston Chou (Netflix), Sana Pandey (MIT CSAIL),
Jonathan Stray (UC Berkeley CHAI), Lo-Hua Yuan (Airbnb\thanks{Affiliation at the time this work was conducted}\ ),
Eytan Bakshy (Meta), Timothy Chan (Statsig), Molly Davies (Pinterest), Maria Dimakopoulou (Uber),
Simon Ejdemyr (Netflix), Kenneth Hung (Meta), Nathan Kallus (Netflix \& Cornell University),
Madhav Kumar (Harvard Business School), Thu Le (Lyft), A. Demetri Pananos (Datadog),
Lee Richardson (Google), Brennan Schaffner (Knight-Georgetown Institute), Rose Tan, Martin Tingley,
Nadia Tomova (Booking.com), Panagiotis Toulis (University of Chicago Booth School of Business),
Wenjing Zheng (Roblox), Zander Arnao (Knight-Georgetown Institute), Dean Eckles (MIT)
}}
\date{}

\begin{document}

\maketitle

\begin{abstract}
Online platforms prioritize long-term business outcomes, yet typical experiments are far too short to measure these outcomes directly. Our goal in this paper is to collect and share industry knowledge on how to make decisions from short-term experiments that are better aligned with long-term outcomes. Based on a daylong workshop with 26 experts from 15 online platforms and 4 universities, we formulate a series of propositions that reflect current industry knowledge.

Participants largely agreed that reversals of sign from short-run to long-run treatment effects are rare, with reversals concentrating in specific cases such as treatments involving content quality signals, hyper-monetization, and pricing. Although the magnitude of treatment effects can shift over time, a ``univariate autosurrogate'', corresponding to the short-run treatment effect on the long-run metric of interest, is often hard to beat. A recurring theme was the importance of surrogates that are not only (or even primarily) unbiased for true long-run outcomes, but that improve decision-making. Thus, participants generally agreed that simple, interpretable surrogates were generally preferable to elaborate but hard-to-explain surrogate indices. Participants also agreed that, due to concerns about confounding and transportability, experimentally-learned surrogates are generally preferable to observationally-learned surrogates. However, the drawback is that learning good surrogates from experiments typically requires a large, representative portfolio of long-run experiments that few platforms possess.

We conclude that there is no substitute for a well-run long-term experiment, whether for learning surrogates or validating them, and we highlight open challenges including evolving treatments, persistent treatments not fully mediated by short-term proxies, and mismatch between experimental samples and the target population.
\end{abstract}

\onehalfspacing

\section*{Introduction}

In this paper, we describe what is known in industry about methods to estimate the long-run treatment
effect of a permanent change to digital services. As noted in Proposition 1.4, platforms prioritize
long-term outcomes that are infeasible to measure in experiments of typical duration. Thus, our aim in
this paper is to collect and share industry knowledge on how to evaluate the effect of a product change
on those long-term outcomes without being able to measure them directly. The goal of the methods we
describe is not necessarily unbiased estimation of long-term treatment effects, although this can be an
important input into decisions, but enabling better decisions that are more aligned with long-term
objectives. Because of this, we consider other factors that contribute to better decision-making by
platforms, such as interpretability.

Our conclusions are based on a daylong workshop with 26 attendees, representing 15 online platforms
and 4 universities. Prior to the workshop, we drafted a set of propositions based on public evidence and
personal experience. The workshop consisted of 8 hours of discussing and refining these propositions.
Subsequently, we worked with participants to refine the propositions and ensure that they reflect current
industry knowledge. We strive to support each proposition with both public evidence and participant
input, though this is not always possible.

The canonical problem we consider is predicting the effect of a change to a ranking algorithm on 6-month
daily active users (DAU), based on 7-day cumulative measured engagement metrics in an A/B test (e.g.,
clicks, likes, comments, reshares, upvotes, return visits). Note that this description is applicable to
recommenders and search engines, and a similar problem may exist for chatbots, where instead of
``ranking algorithm'' we can consider tweaks to the Large Language Model (LLM) or the system prompt.
We use this ``canonical problem'' to help with concreteness and avoid ambiguity; however, many other
related problems are in-scope to this work, e.g., different outcomes, different types of treatment, and
different time horizons. We do not cover methods to estimate the aggregate effects of many changes
(e.g., all the changes to a recommendation system in a 6-month period), whether by using a long-running
holdout experiment or not.

\subsection*{Definitions}

\begin{proptable}
\textbf{Experiment} & A randomized A/B test in which users (unless otherwise specified) are randomly assigned to two or more conditions. \propsep
\textbf{Proxy metric} & An individual short-run metric that may be informative about long-run effects. \propsep
\textbf{Surrogate index} & A prediction of the long-run outcome based on the short-term proxy metric(s) and covariates. In the context of an experiment, the surrogate index provides an estimate of the treatment effect on the long-run outcome as a function of treatment effects on short-run proxies \citep{Athey2026}.

We note that participants also used ``surrogate index'' informally to describe any function of short-run treatment effects that is intended to be interpreted as an estimate of the long-run treatment effect. The simplest example is an ``autosurrogate'': the short-run effect on a metric, possibly multiplied by a conversion factor, as an estimate of its long-run effect. Occasionally, we will also use ``surrogate'' informally to match this usage.

We distinguish two ways of estimating surrogate indices:
\begin{itemize}[nosep]
  \item Observation-based surrogacy. See Propositions 4.1--4.4.
  \item Experiment-based surrogacy. See Propositions 5.4--5.6.
\end{itemize} \propsep
\textbf{Overall evaluation criterion (OEC)} & A single metric, often based on one or more surrogates, that is used to decide the overall success of a treatment. \propsep
\textbf{Short-run effect} & The effect after 7 days or duration of a typical experiment on the users enrolled into the experiment (which varies widely by platform). \propsep
\textbf{Long-run effect} & The treatment effect on the population of users after some longer, context-specific time. In some cases, we can think of this as an asymptotic treatment effect, if treatments stabilize in some short enough period of time.
\end{proptable}

\section{Typical industry experimentation practices}

In this section we describe typical experimentation practices in industry as well as the challenges
platforms face in using these experiments to evaluate product changes.

\begin{proptable}
\textbf{1.1 Platforms with over 10 million monthly active users (MAU) run 10,000--100,000+ experiments per year.} &
\textbf{Public evidence:}
\begin{itemize}[nosep]
  \item Booking.com runs more than 25,000 tests a year \citep{HBR2020}.
  \item Microsoft, Amazon, Booking.com, Facebook, and Google each conduct more than 10,000 online controlled experiments annually \citep{HBR2017}.
  \item Microsoft runs ``$\sim$100k A/B tests annually'' \citep{Bajpai2025}.
  \item Meta runs ``hundreds of thousands of experiments'' per year \citep{Meta2024}.
\end{itemize} \propsep
\textbf{1.2 Typical experiment duration ranges from 3 days to 3 months.} &
Participants agreed that the duration required for typical experiments varies widely, and is heavily dependent on user behavior patterns on the platform. For example, social media platforms may be able to run experiments for less than a week, but travel booking platforms often need to wait much longer to observe changes in user behavior.
\\
& \textbf{Public evidence:}
\begin{itemize}[nosep]
  \item ``Most experiments in the industry run for 2 weeks or less'' across 13 different organizations \citep{Gupta2019}.
  \item Median test length from platforms using third-party experimentation platforms ``typically spans 1 to 3 weeks'' \citep{VWO2023}.
\end{itemize} \propsep
\textbf{1.3 At least 7 platforms run some single-feature experiments lasting at least 3 months.} &
In addition to the platforms listed in the public evidence below, participants listed three more platforms that run some single-feature experiments (i.e., experiments where only a single product feature is changed for the treatment group) for 3+ months.
\\
& \textbf{Public evidence:}
\begin{itemize}[nosep]
  \item Google Search ran an experiment on long-term user outcomes measuring ad blindness and sightedness \citep{Hohnhold2015}.
  \item Facebook ran an experiment on notifications for more than one year \citep{Meta2022}.
  \item Pinterest has run an experiment on badging exceeding one year \citep{Egan2015}.
  \item Airbnb runs experiments on pricing algorithms that typically last 4 months \citep{LeDeng2023}.
\end{itemize} \propsep
\textbf{1.4 Platforms prioritize long-term outcomes that are infeasible to measure in typical experiments.} &
Participants universally agreed that while platforms prioritize long-term business outcomes (usually monitored on a quarterly if not annual basis), it would be infeasible to run experiments long enough to observe those outcomes directly. Instead, the dominant industry practice is to run short-run experiments and to extrapolate the treatment effects observed in those experiments to the long term. \propsep
\textbf{1.5 Estimated effects from experiments are one of several inputs that inform decision-making for platforms.} &
Participants agreed that estimated long-term effects are useful both for understanding and decision-making. Long-term metric effects are not the only input into launch decisions, which will often also depend on judgments of costs and benefits not captured in these aggregate metrics.
\end{proptable}

\section{Typical treatment effect trajectories}

Given typical industry practices, a reasonable concern is whether treatment effects measured in
short-run experiments generally align with treatment effects in long-run experiments. The extent to which
effects can shift over time also has important implications for the use of surrogate indices and the
likelihood of encountering the ``surrogate paradox,'' wherein a short-term surrogate gets the sign of the
long-run treatment effect wrong \citep{Chen2007,VanderWeele2013}.

Here, we present propositions that gained consensus among workshop participants regarding the typical
trajectory of treatment effects in instances where both short-run and long-run observations are available.

\begin{proptable}
\textbf{2.1 Typically the signs of the long-run and short-run treatment effects are the same.} &
Most participants did not regard sign reversals from short-run to long-run treatment effects to be a prevalent concern. Participants agreed that sign reversals are likely to be overstated in the public literature because they are more interesting. There are specific cases where sign reversals are more likely, and we document these in Proposition 2.2.

A lack of sign reversal does not mean that the magnitude of the treatment cannot shift in consequential ways over time (i.e., that would change the decision made based on the experiment). We document magnitude shifts in Propositions 2.3--2.5.
\\
& \textbf{Public evidence:}
\begin{itemize}[nosep]
  \item Google found that a short-run effect is usually a reasonable proxy for the long-run effect, at least directionally \citep{Tripuraneni2024}.
  \item In a meta-analysis of 200 A/B tests at Netflix, researchers report that the launch decision at two weeks would agree with the launch decision at two months in 95\% of experiments \citep{Zhang2024}.
  \item Microsoft noted that they ``could not find a single experiment where a statistically significant result in one direction became statistically significant in the other direction'' \citep{Kohavi2012}.
  \item Google reports that treatment effects in ad load experiments are well-approximated by a simple exponential curve whose magnitude rises steeply in the initial period before stabilizing \citep{Hohnhold2015}.
  \item At Pinterest, a badging treatment sharply increased DAU soon after the experiment. While the magnitude of the treatment effect decreased (from 7\% to 2.5\%) over time, the long-run treatment effect was still positive \citep{Egan2015}.
\end{itemize} \propsep
\textbf{2.2 In specific cases, the long-run and short-run treatment effects can have reversed signs.} &
Participants highlighted three cases where sign reversals are more common:
\begin{itemize}[nosep]
  \item Changes to content quality or relevance signals.
  \item ``Hyper monetization'' in gaming platforms that increases revenue in the short run but hurts it in the long run because of lost user retention.
  \item Pricing experiments where lead-day bias is present.
\end{itemize}
\\
& \textbf{Public evidence:}
\begin{itemize}[nosep]
  \item A previous industry whitepaper found that content quality signals had a negative relationship with predicted engagement, but can increase long-term retention \citep{cunningham2025ranking}.
\end{itemize}
\\
&\vspace{-2\baselineskip}
\begin{itemize}[nosep]
  \item Facebook filtered notifications to only show relevant notifications. They found that time spent initially declined, but after a year had increased relative to control \citep{Meta2022}.
\end{itemize}
\\
&\vspace{-2\baselineskip}
\begin{itemize}[nosep]
  \item Bing intentionally degraded relevance on search and found that ``users spent more effort on refining their queries to complete the same task, but gradually lost confidence so they came back less frequently'' \citep{Song2013}.
\end{itemize}
\\
&\vspace{-2\baselineskip}
\begin{itemize}[nosep]
  \item Facebook assigned a random group of users to have fewer quality terms in their ranking function (``minimal integrity holdout''). The biggest impact came from disabling the ``clickbait'' and ``adsfarm'' quality terms. After 1 month these users had higher overall activity by most metrics, e.g., impressions were up by around 0.4\% \citep{FBArchive2019}. However, after 2 years these users had lower overall activity \citep{cunningham2025ranking}.
\end{itemize}
\\
&\vspace{-2\baselineskip}
\begin{itemize}[nosep]
  \item YouTube reduced the ranking score of videos classified as ``trashy'' or ``tabloid-style.'' After three weeks watch time was down by 0.5\%, but after three months watch time recovered and increased relative to a holdout group \citep{cunningham2025ranking}.
\end{itemize}
\\
&\vspace{-2\baselineskip}
\begin{itemize}[nosep]
  \item Airbnb found that lead-day bias can cause differences in the signs of the short-run vs long-run treatment effects \citep{LeDeng2023}.
\end{itemize} \propsep
\textbf{2.3 Novelty effects are common, but their impacts on decisions are variable.} &
Most participants agreed that user novelty effects are common -- i.e., it is common for the magnitude (not the sign) of a treatment effect to shift after the first few days of an experiment. However, participants disagreed about how much novelty effects would actually change the decision in an experiment. One platform shared that they disregard the first few days' worth of data to avoid novelty effects, whereas some participants felt that novelty effects wouldn't typically change decisions. \propsep
\textbf{2.4 Retention-based metrics (e.g., DAU) are especially prone to shifts in magnitude of treatment effect after typical experiment duration.} &
Participants noted that this is a challenge across a variety of platforms:
\begin{itemize}[nosep]
  \item On social media, changes in content ranking can have an effect on DAU with a ``half-life,'' meaning the time until the short-term effect reaches half of the long-term effect, of around 1 month.
  \item On platforms that involve specified time intervals where users must make subscription or software renewal decisions (e.g., every 30 days), retention can be hard to accurately measure in typical-duration experiments.
\end{itemize}
\\
& \textbf{Public evidence:}
\begin{itemize}[nosep]
  \item A YouTube experiment diversifying recommendations on the home page showed an increase in the number of users who watched at least one video per day. They ran this experiment for more than 4 months and found that ``this effect becomes even more pronounced over time'' \citep{Wilhelm2018}.
\end{itemize} \propsep
\textbf{2.5 Treatment effects for changes in ad load can continue to shift 2+ months into the experiment.} &
\textbf{Public evidence:}
\begin{itemize}[nosep]
  \item In a multi-year ad load experiment, Pandora found that treatment effects after 1 year could be more than double the effects at 2 months \citep{Goli2025}.
  \item Google describes the results of many long-term experiments in mobile ad load, noting that ``the learned effects still increased, even after many weeks of treatment'' \citep{Hohnhold2015}.
\end{itemize}
\end{proptable}

\section{General findings about surrogacy}

In this section we discuss findings about surrogacy generally, regardless of whether the surrogate is
estimated with observational data or experimental data.

\begin{proptable}
\textbf{3.1 It is often hard to beat a univariate autosurrogate.} &
Participants broadly agreed that it is often hard to improve upon a univariate autosurrogate (the short-run version of the outcome of interest). For example, the target outcome may be the average of per-day observations taken over the experiment, in which case the short-run outcome enters directly into the long-term outcome. That being said, participants generally conceptualized performance in terms of decision consistency (e.g., agreement in the sign and/or statistical significance of short-run and long-run treatment effects), which may be a lower bar than bias or RMSE.
\\
& \textbf{Public evidence:}
\begin{itemize}[nosep]
  \item Netflix analyzed such auto-surrogates, finding high consistency between conclusions drawn from auto-surrogates compared with the true long-term outcomes. This creates an upper bound for possible improvements from more elaborate surrogate indices \citep{Zhang2024}.
\end{itemize} \propsep
\textbf{3.2 In specific cases, univariate autosurrogates can perform poorly, so it may be possible to improve on them.} &
Participants flagged the following scenarios as ones where autosurrogates may not perform well relative to more complex surrogate indices:
\begin{itemize}[nosep]
  \item There may not be a straightforward definition of the autosurrogate. For example, supposing the long-run outcome of interest is the proportion of users who are MAU, there is no obvious short-run version of MAU observed in experiments that are run for less than a month.
  \item The target outcome may consist of infrequent, one-time events. Travel bookings are a clear example of an infrequent transaction for which we need to have other signals of intent.
  \item On platforms where there are short-term tradeoffs between engagement and revenue, a revenue autosurrogate may not be a good proxy for long-term revenue.
  \item When launching a new product, or making radical changes to an existing product.
\end{itemize} \propsep
\textbf{3.3 Elaborate surrogate indices can be hard to explain, limiting their usefulness.} &
A common approach to observational surrogacy, which we discuss in detail in Section 4, is to fit a model of the long-term outcome based on short-term proxies to observational data \citep{Athey2026}. Such models constitute valid surrogate indices under strong assumptions, including unconfoundedness. To make these assumptions plausible, participants noted the appeal of using flexible models, possibly with a large number of variables (whether proxies or potential confounders).

However, the fact that these models are harder to interpret and explain limits their usefulness in practice. For example, participants agreed that stakeholders often need to know why a surrogate index moved in a given experiment, which requires an understanding of the importance of each individual proxy in the surrogate. In short, being unable to explain why a surrogate moved erodes trust in the surrogate. As a consequence, participants were generally skeptical of ``black box'' surrogate indices and preferred simpler surrogates, including autosurrogates. \propsep
\textbf{3.4 Estimates of long-run true effects are subject to human judgment.} &
Estimates from surrogate indices will represent true best-guesses only under strict assumptions. For example, in the case of experimental surrogacy, one assumes that the new experiment is a draw from the same pool of experiments. However, in practice, participants noted that often decision-makers know how an experiment differs from past experiments, among other information (e.g., they may know from user experience research that user responses to the treatment are likely to be dynamic), and it is appropriate to make adjustments based on that information.
\end{proptable}

\section{Observational estimation of surrogate indices}

Here, we consider estimation of surrogate indices from observational data. A valid surrogate index
estimated in this way requires the usual assumptions of observational causal inference, such as no
unmeasured confounding between the proxies and the outcomes. Yet it also requires other assumptions,
such as full mediation and transportability from the observational sample to the new experiment \citep{Athey2026}.

Many participants found these assumptions dubious for their platforms. For example, media platforms
questioned whether one could ever eliminate the confounding between short- and long-term
engagement. That being said, here participants also returned to the recurring tension between unbiased
estimation and decision-making: even a biased observational surrogate index can be an informative input
into launch decisions.

\begin{proptable}
\textbf{4.1 There are strong reasons to expect bias in observational surrogacy on platforms.} &
Participants broadly agreed that confounding bias should be expected in observational surrogacy. This is because, as in observational causal inference more generally, we expect that proxies are correlated with long-run outcomes for various reasons that do not reflect causal effects. For example, more active users in the short run are also more likely to engage and be retained in the long run in part due to latent characteristics (e.g., free time).
\\
& \textbf{Public evidence:}
\begin{itemize}[nosep]
  \item In a 21-month Pandora ad load experiment comparing observational methods against experimental treatment effects, a bevy of observational methods (including those leveraging panel data, rich controls and fixed effects, and instrumental variables) often led to biased estimates of long-term effects, at times with the incorrect sign \citep{Goli2025}.
\end{itemize} \propsep
\textbf{4.2 Decisions made with an observational surrogate index, even if biased, can be highly consistent with decisions made using the true long-term outcomes.} &
Although not all organizations had implemented observational surrogate indices, participants from two organizations that did have observational surrogates reported that treatment effects on those surrogates were highly correlated with (if not unbiased for) treatment effects on long-term outcomes in A/B tests. This was evaluated based on the precision/recall of launch decisions--i.e., whether launch decisions based on the surrogate would generally have agreed with launch decisions based on the long-term outcome--rather than accuracy of the predictions.

However, participants noted several caveats to the usefulness of this kind of surrogate:
\begin{itemize}[nosep]
  \item Precision/recall of decisions does not imply that these metrics can provide unbiased estimations of the long-term effect (see issues raised in Propositions 4.1 and 4.3). The former usually implies a different loss function than the latter (e.g., disagreement in sign rather than mean-squared error). Yet, accurate estimates of long-term treatment effects are also often an important ingredient into decision-making.
  \item In practice, due in part to the lack of long-run experiments, validations often have little statistical power to detect violations of key assumptions. This means that platforms cannot always detect when observational surrogates will be misaligned with the true long-run effect.
\end{itemize}
\\
& \textbf{Public evidence:}
\begin{itemize}[nosep]
  \item \citet{Yang2023} find that policy learning based on a surrogate index led to similar results to policy learning based on the ground truth long-term outcomes; however, the test comparing the two results had low power.
\end{itemize} \propsep
\textbf{4.3 Standard statistical inference with observational surrogate indices can underestimate uncertainty.} &
Participants noted that effects estimated with a surrogate index often have much narrower confidence intervals than if using the long-run metrics from an experiment. This is particularly true when the proxies used only capture a small fraction of the variation in the long-run outcome, as this will decrease the standard errors of treatment effects estimated with this (low variance) imputed outcome. Although the resulting confidence intervals reflect ``known unknowns'' (i.e., well-defined statistical uncertainty) when the surrogacy assumptions hold, they do not reflect ``unknown unknowns'' (i.e., residual confounding bias).

The resulting precision can be a double-edged sword. On the one hand, precise estimates are useful for making faster and more confident decisions. On the other hand, a precise estimate that is later invalidated can greatly erode trust. \propsep
\textbf{4.4 When possible, experimental surrogacy is preferable to observational surrogacy.} &
Given the various limitations of observational surrogacy described above, participants were in broad agreement that whenever possible, it is better to use experimental surrogacy than observational surrogacy.
\end{proptable}

\section{Using experiments to evaluate and estimate surrogate indices}

Given the clear threats to the credibility of surrogate indices learned from observational data, it is natural
to look to long-run experiments to learn, or at least evaluate, surrogate indices. Nonetheless, difficult
statistical and practical challenges remain.

\begin{proptable}
\textbf{5.1 One cannot validate a surrogate index without running at least some long-run experiments, whether of individual features or team- or product-level holdouts.} &
Participants broadly agreed that platforms can get immense learnings from even a few long-run experiments. This is because a single (or a few) long-run experiments can be used to \emph{evaluate} how well a given surrogate index can forecast long-term treatment effects, or how well a decision rule based on a surrogate performs -- although more experiments are obviously needed to \emph{learn} surrogate indices solely using experimental variation (see Propositions 5.4--5.6 below).
\\
&
To be sure, the returns to running long-run experiments need to be carefully weighed against the costs (e.g., engineering complexity and maintenance, the opportunity cost of withholding a better experience from some members).
\\
&
However, participants agreed that the value of being able to validate a surrogate index is often worth the cost of at least some long-run experiments. \propsep
\textbf{5.2 Not every long-run experiment is equally useful for all surrogate indices.} &
Participants generally agreed that surrogate indices only perform well among experiments of the same class. For instance, applying a surrogate index from a notification experiment to a ranking experiment could yield very misleading results. Similarly, surrogate indices learned from older notification experiments can become outdated as the nature of those experiments evolves over time.
\\
&
An especially poignant example raised by one participant is ``metric gaming'': one might find a correlation between a short-term outcome and the long-term north star (for example, engagement at 7 days and at 60 days) in a class of past experiments, then try to move that short-term outcome in ways that break the historical causal structure (for example, by sending notifications every 7 days).
\\
&
Alternatively, at least one participant opined that more broadly-defined surrogates (e.g., an OEC) can be applicable among experiments from different classes. A useful way to reconcile these perspectives is to take a mechanistic view of surrogates as treatment mediators \citep{Bibaut2024}. This can help to inform which experiments should be compared (i.e., those that affect the same mechanisms) versus not.
\\
&
Another potential complication is that many long-run experiments are ``stacked'' with multiple treatments, with individual treatments being added over time. The resulting experiment affects multiple mechanisms across multiple time windows, further challenging interpretation (see Proposition 6.3 below). \propsep
\textbf{5.3 Even a portfolio of long-run experiments may not be sufficient for evaluating and/or learning surrogates.} &
Even if organizations have a portfolio of long-run experiments, often there is substantial selection into this portfolio, raising concerns about generalizability:
\begin{itemize}[nosep]
  \item Teams typically end run-of-the-mill experiments with very large impacts, whether positive or negative--so long-running experiments tend not to have dramatically large effects.
  \item While some platforms run dedicated long-term holdouts for learning and/or progress tracking, due to the high operational and engineering burden, these long-term holdouts are typically reserved for the most impactful treatments.
  \item There may be other subtle biases, relating to the use of expert judgment, that affect which experiments are continued --- which may rely on signals not in the surrogate.
\end{itemize}

Participants noted multiple challenges that this sample selection introduces:
\begin{itemize}[nosep]
  \item Positivity/overlap problems, whereby some short-run effect estimates will be far outside the support of the long-run experiments used for training.
  \item Asymmetry in evidence: since platforms are especially likely to end obviously bad treatments early, we have more data points to validate positive treatment effect estimations vs. negative effect estimations, and we inherently make the assumption that surrogate quality is symmetric.
\end{itemize}

\propsep

\textbf{5.4 Many organizations have too few relevant long-run experiments to effectively learn surrogate indices solely using experimental variation.} &
In addition to evaluation, long-term experiments can help organizations learn effective surrogates. However, in order to estimate a surrogate index solely from experimental variation, one needs to have at least as many long-run experiments as components in the surrogate index. Yet, for reasons discussed under Proposition 5.3, the number of suitable long-term experiments is often small. Additionally, treatment effects on the short-term proxies may be small, which drives up statistical variance and induces asymptotic bias related to the weak instruments problem \citep{Bibaut2024}. Even if each experiment is large, without many long-term experiments, estimates can be imprecise (or even unobtainable without regularization). \propsep
\textbf{5.5 Naively regressing estimated treatment effects on long-run outcomes on estimated treatment effects on short-term proxies will yield a biased estimate of the effects of the proxies on the outcomes.} &
Even if one has a large number of experiments, if the treatment effects on the short-term proxies are weak, then regression of the estimated long-term treatment effects on the estimated short-term proxy treatment effects will be asymptotically biased. The root cause of this bias is correlated measurement error: treatment arms that have spuriously high values of the proxies will also have spuriously high values of the long-term outcome if these are positively correlated in observational data \citep{CunninghamKim2022}.

The bias can be illustrated by plotting placebo experiments, generated by randomly reshuffling the treatment assignment. Although there are no true treatment effects by construction, we will ``find'' strong between-experiment relationships across any pair of metrics, which is driven entirely by correlated measurement error. \propsep
\textbf{5.6 Various techniques are available to reduce bias from correlated measurement error, though many require more than experiment-level statistics.} &
These techniques include:
\begin{enumerate}[nosep]
  \item Running bigger experiments. Since the impact of measurement error shrinks with sample size, a ``simple'' design-based remedy is just to allocate more users to each experiment \citep{Zito2025}. However, in practice platforms are often limited in their ability to run larger experiments, for example due to contention for traffic between experiments.
  \item Running a regression just using the high signal-to-noise metrics. Dropping noisy proxies can help with estimability and interpretability of regressions of long-term treatment effects on short-term treatment effects, but at the cost of bias if the proxy is an important mediator, so it's hard to know when this will be a good idea without an explicit model. See \citet{Bibaut2024} for a structural interpretation of this regression.
  \item Running a regression just using the strongest \emph{experiments}. The optimal proxy is dependent on the size of the experiment: in larger experiments, one can afford to rely on less sensitive but better aligned proxies \citep{Tripuraneni2024,Chou2026}. If the distribution of experiments is fat-tailed, then the strongest experiments will have higher signal-to-noise ratio, and so lower bias. This can motivate the use of L0 regularization in the treatment effects-on-treatment effects regression, for example \citep{PeysakhovichEckles2018}. The drawback with this approach is that it estimates the relationship using only outliers, making nonlinearity a greater concern.
\end{enumerate}
\\
&
\begin{enumerate}[nosep,start=4]
  \item Adjusting for the bias. The measurement error can usually be estimated and adjusted for explicitly. Netflix provided various estimators drawn from the weak instruments literature that incorporate estimates of the measurement error \citep{Bibaut2024}. See also \citet{CunninghamKim2022} and \citet{Tripuraneni2024} for an empirical Bayesian approach to this problem. Meta uses experiments as instruments for building a (regularized) predictive model of long-term changes for new experiments \citep{Day2026}.
  \item Using experiment splitting. Each experiment can be randomly divided into one or more folds. This effectively turns a set of experiments into a set of pairs (or triads, etc.) of experiments, each of which has experiments with identical treatment effects but independent noise. Thus, one can run a regression with the left-hand sum from one split, and the right-hand sum from the other split, eliminating the effect of correlated measurement error. This can be extended to nonlinear models \citep{CoeyCunningham2019} and general nonparametric function approximators \citep{vanderLaan2025a}, though in practice the usual caveat applies that decision-makers generally prefer interpretable surrogates.
  \item Backtesting and validating decision rules, causal structure aside. One can also sidestep structural assumptions and instead try to find proxies that, if acted upon, optimize the long-term metric. Netflix shows how cross-validating decision rules, such as ones based on surrogate indices, yields consistent selection of such rules \citep{Chou2025}.
\end{enumerate}
\end{proptable}

\section{Open challenges in estimating long-term effects from short-term experiments}

We conclude by listing open challenges that gained consensus from participants as relatively
understudied and/or unaddressed. By definition, these propositions have less public-facing evidence;
however, we point to relevant literature where available and call for future research on these challenges.

\begin{proptable}
\textbf{6.1 Systematic evidence on treatment effect trajectories is scarce.} &
Although participants generally agreed that treatment effect reversals are rare (Proposition 2.1), evidence of this that is both systematically collected and public-facing remains scarce.

A major contributing factor is the high cost of running long-term experiments. Furthermore, due to this cost, long-term experiments tend to be limited to the most significant changes, introducing selection bias. Therefore, an important challenge for future research is simply to catalog treatment effect trajectories -- of both major changes and more typical changes -- more systematically. \propsep
\textbf{6.2 The reasons why treatment effects change over time are likely varied, and it may be possible to make surrogates more credible by specifying why short-term treatment effects differ from long-term effects.} &
Participants generally agreed that there are many reasons why treatment effects could evolve over time---including calendar effects and seasonality; novelty and primacy effects; pull-forward effects; and the exhaustion of high-quality results---and that understanding these mechanisms is helpful for reasoning about the bias of short-term effects (see examples in Proposition 2.2).

This suggests that it may be possible to design more credible surrogates by explicitly accounting for these mechanisms, such as in Airbnb's handling of lead-day bias in pricing experiments \citep{LeDeng2023}. One tradeoff here is that these mechanisms may be highly platform-dependent, limiting the generality of surrogate methods that address them.

Future research should probe the drivers of time-varying treatment effects more systematically. In particular, different mechanisms may be responsible for attenuation, amplification, and/or reversals in sign of the treatment effects. \propsep
\textbf{6.3 Many long-term experiments evaluate an evolving treatment rather than a single fixed intervention. As the treatment is iterated upon and updated, the treatment effect reflects the impact of a sequence of interventions.} &
Although long-term holdbacks are the gold standard for evaluating long-term effects, in practice, these may be structured as cumulative holdbacks where one treatment arm is continually updated with new treatments. For example, very large changes (e.g., a new UI) may be tested as a long-running experiment in which the experience is continuously refined over time.

Even ignoring active iteration on features, algorithms can evolve according to the training data they collect. For example, a contextual bandit model that trains on both production and randomized data will evolve even without active development. Attempts to segregate data accessed by adaptive algorithms may actually create differences between experimental data and after launch \citep{Si2024}.

One way to frame this is to define longitudinal potential outcomes as a function of a \emph{sequence} of treatments, $Y_{i,t}(Z_1, Z_2, \ldots, Z_t)$. Estimands of interest then become contrasts between treatment sequences over time \citep{Shi2025}.

In such long-term experiments, some useful questions for future research to explore are:
\begin{itemize}[nosep]
  \item How do we infer the long-term effect of the eventual treatment after all refinements?
  \item Is it possible to isolate the long-term effects of individual refinements, possibly with the aid of short-term experiments of only those refinements?
  \item Is it possible to identify key interactions between refinements?
\end{itemize} \propsep
\textbf{6.4 Although many surrogacy methods assume a point-in-time treatment whose effects are fully mediated by short-term outcomes, other methods may be necessary to model the effect of a persistent treatment (such as a permanent change to an algorithm).} &
By definition, a permanent change to a digital service means that the treatment continues to affect long-term outcomes after the experiment, violating the classical surrogacy assumptions. However, under certain structures (e.g., a linear structural model or the InSIDE assumption in Mendelian randomization \citep{Sanderson2022}), one can still credibly talk about estimating the effect mediated by observed surrogates \citep{Bibaut2024,Shi2025}. Another generalizable structure that allows us to project out short-term treatment effects is Markov structure \citep{Tran2024,vanderLaan2025b}.

In addition to surrogacy methods, future research should explore these and other alternatives to assuming that the long-term effect of treatment is fully mediated through short-term outcomes. \propsep
\textbf{6.5 The experimental sample may differ systematically from the target long-term population.} &
Short-term experiments can oversample on highly active users (``activity bias'') and, in the extreme, fail to include any users who are not active on the platform during the test window. The treatment effects on such users may be systematically different, making the true long-term effect unidentified without other assumptions.

Secular population drift (e.g., acquisition of new users) can also induce systematic differences between experimental samples and the true long-term population.

Future research -- for example, into the intersection of surrogacy and transportability \citep{PearlBareinboim2011} -- should explore how to adjust long-term treatment effect estimates for factors like activity bias and acquisition, which are often relevant to business stakeholders. \propsep
\textbf{6.6 Projections of long-term effects are only valuable to platforms insofar as they enable better decisions, which can entail unresolved tensions between simplicity and accuracy.} &
Participants agreed that simplicity of the surrogate index is valuable for building trust with decision-makers. At the same time, inaccuracy of the surrogate can obviously erode trust. Thus, two areas for future research are, first, how to balance transparency and accuracy when extrapolating from short- to long-term treatment effects, and second, how to propagate error appropriately, whether statistical or deterministic.
\end{proptable}

\section*{Conclusion}

This workshop was a unique opportunity to share practical wins and experience from attempts to
estimate the true long-run treatment effect of changes to digital services.

As our paper shows, there are many open questions and areas of disagreement (see, for example,
Propositions 2.3, 4.2, and 6.1--6.6). Because so many of our findings are highly context-dependent (e.g.,
on the type of platform), we recommend readers rely on individual propositions along with their own
judgment and context when attempting to evaluate for long-term outcomes.

However, we did agree on two broadly-applicable recommendations. First, longer-running experiments
are immensely valuable in evaluating a surrogate, even if a platform can only run a few of these
experiments and even if they aren't optimally designed to evaluate a surrogate. That being said, not all
long-term experiments are equally useful for all surrogates and care must go into how they are designed
and interpreted. Second, due to the risk of confounding and nontransportability, experimentally-learned
surrogates are generally preferable to observationally-learned surrogates. However, this assumes that a
reasonably large and representative portfolio of long-run experiments is available. When such a portfolio
is not available, one must naturally rely on observational surrogates or simple autosurrogates, which can
yield highly consistent decisions even if biased. Even still, it bears repeating that, when it comes to
evaluating surrogates for long-term outcomes, there is simply no substitute for a well-run long-term
experiment.

\section*{Acknowledgments}

We would like to thank the MIT Initiative on the Digital Economy for helping us make the paper possible.

We would also like to thank James McQueen for his help running the workshop and for his substantive
contributions to the propositions.

This material is based upon work supported by the National Science Foundation Graduate Research
Fellowship Program under Grant No. (2141064). Any opinions, findings, and conclusions or
recommendations expressed in this material are those of the author(s) and do not necessarily reflect the
views of the National Science Foundation.

\bibliographystyle{apalike}
\bibliography{references}

\end{document}